\documentclass[usenatbib]{mnras}
\usepackage[T1]{fontenc}
\usepackage[utf8]{inputenc}
\usepackage{blindtext}
\usepackage{comment}
\usepackage{amsmath}
\usepackage[dvipsnames]{xcolor}
\usepackage{array}
\usepackage{graphicx}
\usepackage{tikz}
\usepackage{amssymb}
\newcolumntype{P}[1]{>{\centering\arraybackslash}p{#1}}

\title[Non-thermal Emission in MBHBs]{Non-thermal X-ray Emission from Merging Massive Black Hole Binaries}

\author[Krauth \& Davelaar]{
Luke Major Krauth$^{1}$\thanks{E-mail: L.M.KRAUTH@uva.nl}
and Jordy Davelaar$^{2,3}$\\
$^{1}$Gravitation Astroparticle Physics Amsterdam (GRAPPA), University of Amsterdam,\\
Science Park 904, 1098 XH Amsterdam, The Netherlands\\
$^{2}$Department of Astrophysical Sciences, Peyton Hall, Princeton University, Princeton, NJ 08544, USA\\
$^{3}$NASA Hubble Fellowship Program, Einstein Fellow
}

\date{Accepted XXX. Received YYY; in original form ZZZ}
\pubyear{2026}

\begin{document}
\maketitle

\begin{abstract}
\normalsize
Recent hydrodynamical simulations have identified a disappearing thermal X-ray signature in massive black hole binaries (MBHBs) embedded in circumbinary disks, arising from the tidal truncation and depletion of minidiscs shortly before merger. This feature has been proposed as a promising electromagnetic counterpart to MBHB mergers detectable by LISA. In this work, we examine whether non-thermal X-ray emission powered by magnetic reconnection could obscure or modify this thermal X-ray drop. We construct semi-analytic models for both the thermal X-ray emission from minidiscs and the non-thermal synchrotron emission produced by reconnection in magnetically dominated black hole magnetospheres. Evaluating these models across the MBHB mass range relevant for LISA, we find that for physically motivated magnetic field strengths and accretion rates, the non-thermal X-ray luminosity remains several orders of magnitude below the thermal component throughout the inspiral, particularly in the soft X-ray band where the thermal emission is concentrated. Even under optimistic assumptions that enhance the non-thermal emission, it remains insufficient to erase the characteristic thermal drop, though a transient hard X-ray enhancement may arise near merger. We further incorporate the magnetospheric balding framework to model the decay of non-thermal emission near merger, finding that reconnection-powered X-ray emission fades on short, mass-scaled timescales once the external magnetic flux supply is disrupted. Taken together, our results indicate that non-thermal emission is unlikely to mask the disappearing thermal X-ray signature, reinforcing its robustness as an electromagnetic counterpart to MBHB mergers and its potential utility for multi-messenger studies with LISA.

\end{abstract}


\section{Introduction}
The cores of most galaxies are thought to host a massive black hole (MBH) \citep{LyndenBell1969}. When galaxies merge, it is expected that the MBHs in each will slowly sink towards each other, eventually becoming gravitationally bound, forming a massive black hole binary (MBHB) \citep{Begelman1980}.

The gas around this MBHB is expected to turn into a thin disk supported by angular momentum \citep{Prendergast1968,Pringle1972,Shakura1973}. Numerical simulations over the last decades have commonly found similar aspects of these MBHB+CBD systems. As the binary orbits, a low density, elongated central cavity is carved into the CBD \citep{Artymowicz1996,MacFadyen2008,Shi2012,Ragusa2016}. Additionally, material is tidally stripped from the closest edge of the cavity wall, and forms into streams that feed the BHs. Smaller discs, commonly called ``minidiscs'' can form around each BH from some of this infalling gas. Meanwhile, other portions of the stream are flung out towards the far end of the cavity and can build into a non-axisymmetric density commonly called the ``lump.'' Once every several binary orbits this lump itself leads to flow patterns around the cavity wall which ultimately modulate the accretion onto the binary \citep{Shi2012,Dorazio2013,Noble2021,Krauth2023,Krauth2025}. Any of these regions, streams, CBD, lump, minidiscs, as well as their interplay, can lead to observable signatures \citep{Westernacher-Schneider2022}.

Amongst this chaos, \cite{Krauth2023} recently found a promising EM signature---a disappearing thermal X-ray luminosity that presents in many MBHB systems just before merger. Using 2D hydrodynamical simulations at high resolution, \cite{Krauth2023} observed that within these MBHB+CBD systems, the minidiscs surrounding each BH are responsible for the most energetic photon emission, which falls primarily in the X-ray band. In the late, GW-dominated inspiral, the minidiscs become tidally truncated, leading to reduced surface area and a corresponding reduction in both their accretion rates and luminosity. As the system approaches merger, in the last hours/days, this loss is so great that there is a several-order of magnitude drop in the thermal X-ray luminosity.

The accretion rate drop responsible for the X-rays' disappearance was also found in isothermal gas models \citep{Dittman2023}. Additionally, the X-ray drop was confirmed for 3D hyper-Lagrangian resolution simulations with up to 2.5th-order Post-Newtonian (PN) corrections \citep{Franchini2024}, intermediate mass ratio inspirals (IMRIs) \citep{Clyburn2025}, for certain models which depended on binary torque \citep{Zrake2025}, and unequal-mass binaries with  0.1$\leq q=M_2/M_1\leq$ 0.5 \citep{Krauth2025}.

Not only could this distinct signature lead to sky localization of host galaxies of MBHB mergers in the LISA error volume, but it could also open new avenues for scientific results. From gravitational waves of these systems, observations can provide detailed information about the component masses, orbital parameters, spin characteristics, and luminosity distance of the MBHB \citep{Bogdanovic2022}. The EM counterparts to the systems can provide us information on the spectral energy distribution (SED), periodic variability, jet orientation, and also orbital parameters \citep{Baker2019,Bogdanovic2022}. When combining this information, one can constrain the speed of gravitational waves \citep{Haiman2009}, test General Relativity against alternative theories \citep{Rham2018, Hassan2012}, provide a relation between merging MBHs and their host galaxies as a function of redshift, luminosity and other properties, allowing us to better understand the co-evolution of MBHs with their host galaxies \citep{Kormendy2013}, as well as gain insight into the primary mechanisms driving the mass growth of MBHs over cosmic epochs (see, e.g., \citealt{Baker2019,Bogdanovic2022} for reviews and references). As such, understanding the potential EM signatures originating from the CBD gas is paramount for these scientific goals.

While many studies have now confirmed this X-ray drop, these works have focused on examining the \textit{thermal} X-ray emission. X-ray emission can however also originate from non-thermal sources. One source potentially relevant to the X-ray drop is that which originates from magnetic reconnection.

When the magnetic pressure near the horizon of a black hole becomes comparable to the ram pressure of the inflowing gas, the accretion flow is expected to enter a magnetically arrested disk (MAD) state \citep{Narayan2003,Tchekhovskoy2011}. In this regime, strong magnetic flux accumulates near the horizon, inhibiting further inflow and promoting magnetic reconnection in the inner magnetosphere.

During the very late inspiral and early post-merger of a MBHB+CBD system, the progressive depletion of the minidiscs and accompanying decline in accretion rate may naturally drive the system toward such a magnetically dominated state \citep{Most2024,Manikantan2025}. Global GRMHD simulations of MADs show that strongly magnetized flows generically develop thin current sheets near the black holes, where magnetic reconnection can efficiently convert magnetic energy into non-thermal particle acceleration and high-energy radiation \citep{Ripperda2020,Chashkina2021,Ripperda2022}. This process is capable of producing luminous synchrotron emission extending into the X-ray and soft-$\gamma$-ray bands.

As the system evolves through merger and the inflow of magnetic flux diminishes, reconnection can rapidly dissipate the large-scale magnetic fields threading the black holes, leading to a ``balding'' of their magnetospheres \citep{Lyutikov2011a,Lyutikov2011b,Bransgrove2021}. The associated high-energy emission would then decay on short timescales, potentially producing a brief non-thermal flare followed by rapid fading.

These considerations suggest that a merging MBHB+CBD system may exhibit a short-lived burst of non-thermal X-ray or soft-$\gamma$ emission coincident with the disappearance of the thermal X-ray signal from the minidiscs. Whether such a non-thermal flare can temporarily obscure or modify the thermal X-ray drop identified by \cite{Krauth2023} remains an open question. Addressing this possibility is the primary goal of this work.

Our main result suggests that, across all realistically feasible regions of parameter space, the non-thermal X-ray component from synchrotron radiation remains several orders of magnitude below the thermal emission throughout the inspiral. Even when time-dependent spectral evolution and band-specific hard X-ray enhancement is taken
into account, the non-thermal contribution strengthens primarily at higher photon energies and only after the thermal luminosity has already declined substantially. Consequently, the disappearing thermal X-ray signature persists as a robust and distinguishable feature, reinforcing its viability as a means for identifying host galaxies of MBHB mergers.

The remainder of this paper is organized as follows. In \S~\ref{sec:Setup}, we outline our semi-analytic models for thermal and non-thermal X-ray emission, including the treatment of magnetospheric balding. In \S~\ref{sec:Results}, we compare the resulting luminosity scalings and decay timescales across the LISA mass range. In \S~\ref{sec:Con}, we summarize our conclusions and discuss observational implications.

\section{Thermal, Non-thermal, and Balding Modelling}
\label{sec:Setup}

\subsection{Thermal X-ray Emission from Minidiscs}
\label{subsec:thermal}

We compute the total thermal X-ray luminosity by constructing a smooth, semi-analytic model spectrum whose shape and normalization are informed by the hydrodynamical simulations of \cite{Krauth2023}. The simulated spectra motivate both the characteristic thermal peak and the overall spectral shape of the emission originating from the minidiscs. We then scale this spectrum with the total binary mass, allowing the peak energy and amplitude to shift according to physically motivated mass dependences consistent with thin-disk emission, while optionally permitting modest spin-dependent corrections to both the radiative efficiency and the effective inner disc radius.

For each mass, the resulting spectral energy distribution $\nu L_\nu$ is converted to $L_\nu$ and integrated over the X-ray band to obtain the total thermal X-ray luminosity,
\begin{equation}
    L_X = \int_{E_{\rm min}}^{E_{\rm max}} L_E \, dE ,
\end{equation}
where we adopt an X-ray energy range of
$E_{\rm X\text{-}ray} = 0.124$--$124.0~\mathrm{keV}$.
This procedure preserves the spectral trends seen in the simulations while enabling a controlled extrapolation across the LISA MBHB mass range, providing the thermal X-ray baseline against which we compare non-thermal emission.

\subsection{Non-thermal synchrotron model from magnetic reconnection}
\label{subsec:nonthermal}

To model the non-thermal X-ray emission powered by magnetic reconnection, we adopt a semi-analytic synchrotron framework calibrated to reproduce the reconnecting-current-sheet spectra shown in Fig.~2 of \citet{Hakobyan2023}. In this picture, relativistic electrons and positrons accelerated during reconnection are described by a power-law distribution with an exponential cutoff,
\begin{equation}
    N(\gamma) \propto \gamma^{-p}\exp(-\gamma/\gamma_c),
\end{equation}
where the spectral index $p$ and cutoff Lorentz factor $\gamma_c$ are chosen to match the kinetic simulation results in the radiative reconnection regime. The synchrotron emissivity is computed using the standard angle-averaged synchrotron kernel, integrating over particle energy and pitch angle to obtain the spectral energy distribution $\nu L_\nu$.

Radiative losses are incorporated following the reconnection-powered synchrotron model of \citet{Hakobyan2023}, in which synchrotron cooling limits the characteristic emission frequency through the synchrotron burnoff condition. In practice, this is implemented by allowing the cooling constraint to enter only through the effective critical synchrotron frequency $\nu_c(\gamma,B)$, while leaving the overall emissivity amplitude undamped. This treatment captures the key result from first-principles kinetic simulations that particle acceleration at reconnection X-points remains efficient even in the strong-cooling regime, while the emitted spectrum develops a high-energy cutoff that is largely insensitive to the magnetic field strength.

For a given magnetic field $B$, the resulting synchrotron spectrum is normalized such that the total radiated power equals a fixed fraction of the available jet power. We parameterize this normalization by assuming that a fraction
\begin{equation}
    L_{\rm rec} = f_{\rm rec} L_{\rm BZ}
\end{equation}
of the Blandford--Znajek jet power $L_{\rm BZ}$ is dissipated through magnetic reconnection and radiated, with $f_{\rm rec}$ taken to be an order-unity efficiency consistent with kinetic reconnection studies and the finding that radiative reconnection converts an $\mathcal{O}(10\%)$ fraction of the Poynting flux into radiation \citep{Sironi2014,Hakobyan2023}. The corresponding non-thermal X-ray luminosity is obtained by integrating the spectrum over a specified energy range. In addition to the full X-ray band considered for our baseline comparison, we evaluate the synchrotron contribution across instrument-motivated soft and hard X-ray sub-bands to assess its observational detectability. We evaluate the reconnection-powered spectrum across the MBHB mass range by computing $B(M,\dot{M})$ and the corresponding $L_{\rm BZ}$ at each point in a $(M,\dot{M})$ grid.

Because the synchrotron spectrum is constructed explicitly as a function of magnetic field strength and accretion rate, this framework allows us to compute not only band-integrated luminosities but also time-dependent spectra and band-limited light curves as the binary separation decreases toward merger. In practice, we implement this through a smooth, separation-dependent scaling of the available reconnection power, motivated by studies that find increasing magnetic stress and dissipation as compact binaries approach coalescence (e.g., \citealt{Hakobyan2023,Manikantan2025}) and by the rapid decline in minidisc accretion rates seen in our simulations during the final stages of inspiral. We emphasize that this prescription is intended to bracket physically plausible behaviour rather than to provide a first-principles prediction of the exact merger light curve or detailed spectral evolution; our goal is to test whether any reasonable enhancement of reconnection-powered emission could plausibly obscure the thermal X-ray disappearance.

We do not assume a fixed reconnecting volume (e.g. $R_{\rm ISCO}^3)$; instead, we normalize the non-thermal component by an energetic budget $L_{\rm rec}=f_{\rm rec}L_{\rm BZ}$, where the relevant near-horizon scale is already implicit in the Blandford--Znajek power. At larger jet scales, the declining magnetic field strength with radius reduces the available magnetic energy density, shifting dissipation to spatially extended regions and favouring broadband, lower-energy emission. We therefore do not explicitly model non-thermal emission from jet--jet interactions during merger, which are expected to contribute as a slowly varying background rather than to the compact X-ray luminosity relevant for assessing the robustness of the thermal X-ray drop \citep{Gutierrez2024}.

In magnetically arrested accretion flows, the near-horizon magnetic field strength $B$ can be related to the dimensionless horizon magnetic flux $\phi_B$; however, we retain $B$ as an explicit parameter here to remain consistent with the calibrated reconnection--synchrotron framework adopted from kinetic simulations. Magnetic field strengths and accretion rates are drawn from physically motivated ranges expected for MAD-like accretion flows, and variations around a fiducial model are used to bracket plausible extremes. The explicit equations and fiducial parameter choices entering this procedure are summarized in Appendix~\ref{app:synch_model}.

\subsection{Magnetospheric balding and decay of non-thermal emission}
\label{subsec:balding}

To model the decay of the non-thermal X-ray emission once the external magnetic flux supply diminishes, we adopt the magnetospheric ``balding'' framework \citep{Lyutikov2011a,Lyutikov2011b}, implementing the decay timescales following \citet{Bransgrove2021}. In this picture, magnetic flux threading the black hole event horizon is removed through plasmoid-mediated magnetic reconnection once the external supply of flux diminishes. General-relativistic kinetic and resistive MHD simulations show that the horizon-threading magnetic flux decays quasi-exponentially in time, with a characteristic e-folding timescale set by the reconnection rate in the equatorial current sheet \citep{Bransgrove2021,Ripperda2020,Ripperda2022}.

In kinetic (collisionless) plasma simulations, appropriate for highly magnetized black hole magnetospheres, the decay timescale is found to be $\tau \sim 100\,r_g/c$, while resistive MHD simulations yield somewhat longer timescales, $\tau \sim 300$--$500\,r_g/c$ (see Fig.~3 of \citealt{Bransgrove2021}). Importantly, these decay timescales are largely independent of the absolute magnetic field strength, provided the plasma remains in the highly magnetized regime, and instead scale linearly with the gravitational radius of the black hole.

We therefore model the fading of the reconnection-powered synchrotron emission as an exponential decay,
\begin{equation}
    L_X^{\mathrm{nth}}(t) \propto \exp(-t/\tau),
\end{equation}
with $\tau$ parametrized as a multiple of $r_g/c$. Applying this prescription across the MBHB mass range yields a mass-dependent decay time for the non-thermal X-ray signal, allowing us to determine how rapidly the reconnection-powered emission fades relative to the thermal X-ray drop from the minidiscs. This provides a physically motivated estimate of whether non-thermal emission can temporally overlap with, or briefly obscure, the disappearing thermal X-ray signature identified by \citet{Krauth2023}.

\section{Results}
\label{sec:Results}

\begin{figure*}
    \centering
    \includegraphics[width=0.9\textwidth]{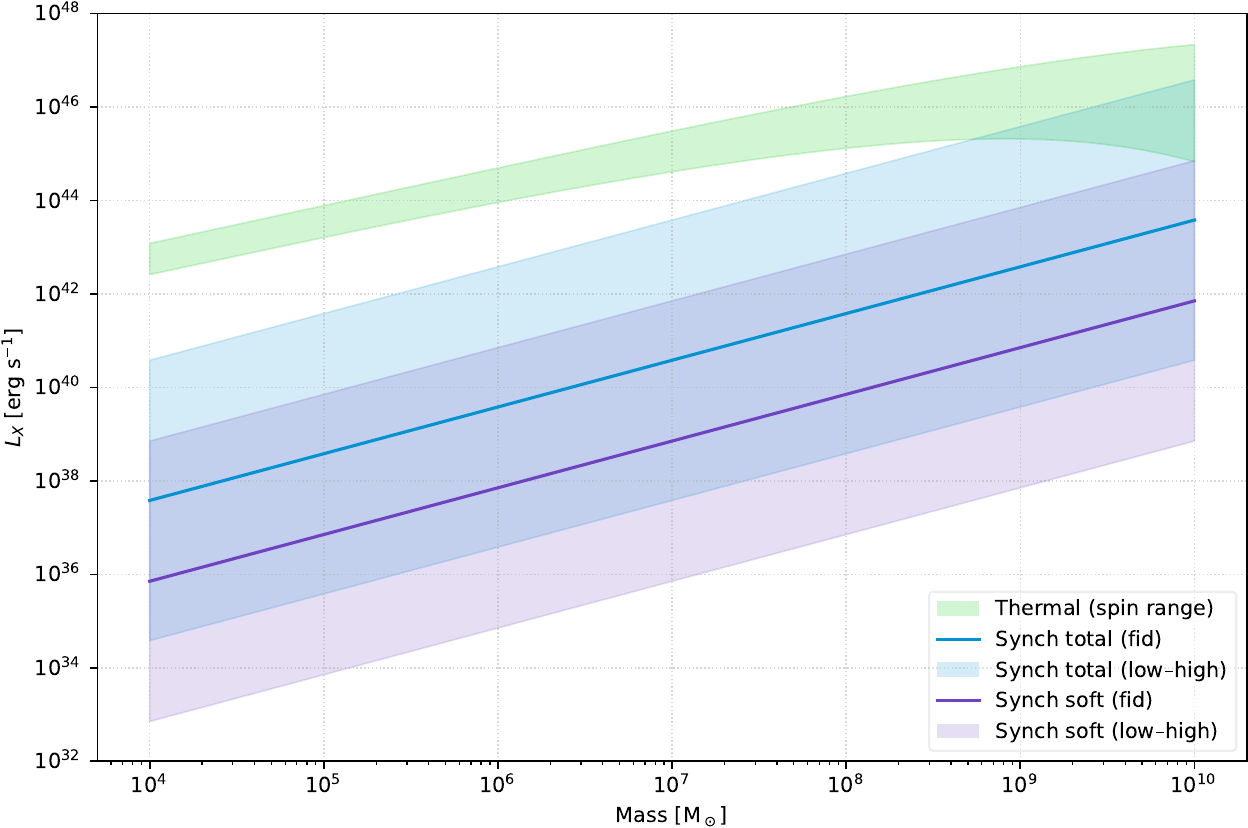}
    \caption{X-ray luminosity $L_X$ as a function of black hole mass. The upper shaded band (green) shows the range of thermal minidisc emission obtained by varying the black hole spin between $a=0$ and $a=0.99$, integrated over the full X-ray band (0.124--124\,keV). The middle shaded band corresponds to the synchrotron luminosity powered by magnetic reconnection integrated over the full X-ray band (``Synch total''), while the lowest shaded band shows the synchrotron contribution restricted to the soft X-ray band (0.124--12\,keV). In each synchrotron case, the solid curve denotes the fiducial model, and the surrounding shaded region spans the range between the ``low'' and ``high'' magnetic-field and accretion configurations. These correspond to $B_0 = 10^{4},\,10^{5},\,10^{6}\,\mathrm{G}$ and $\dot{M} = 10^{-3},\,10^{-2},\,10^{-1}\dot{M}_{\mathrm{Edd}}$, respectively, using the magnetic-field scaling of Eq.~\eqref{eq:A1}. Overall, the thermal minidisc emission remains dominant across the LISA-relevant mass range for physically motivated parameters.
}
    \label{fig:compare_lx_sweeps}
\end{figure*}

Figure~\ref{fig:compare_lx_sweeps} compares the mass scaling of the thermal X-ray luminosity from minidiscs to the non-thermal synchrotron contribution powered by magnetic reconnection. The upper shaded band shows the thermal emission integrated over the full X-ray range (0.124--124\,keV), spanning the variation produced by black hole spin between $a=0$ and $a=0.99$. The two lower shaded bands show the reconnection-powered synchrotron luminosity, evaluated both over the full X-ray band (``Synch total'') and restricted to the soft X-ray range (0.124--12\,keV). In each synchrotron case, the solid line indicates the fiducial model, while the surrounding shaded region brackets the ``low'' and ``high'' magnetic-field and accretion configurations described in Appendix~\ref{app:synch_model}. Although the thermal spectrum is integrated over the full X-ray band for consistency, the emission is effectively confined below $\sim 12$~keV across the mass range considered. This makes the soft X-ray comparison particularly informative, since the full-band thermal luminosity is nearly identical to its soft-band contribution and most current and planned X-ray observatories operate over limited energy sub-bands rather than the entire X-ray range. At the highest black hole masses, the thermal band exhibits a downturn; this does not indicate an intrinsic suppression of minidisc emission, but instead reflects the shift of the thermal spectral peak to lower energies as $M$ increases, causing an increasing fraction of the emission to fall below the adopted X-ray bandpass.

For fiducial, physically motivated magnetic field strengths and accretion rates, the non-thermal synchrotron component remains well below the thermal X-ray emission across the LISA-relevant mass range. In the soft X-ray band, the synchrotron contribution remains subdominant even in the most optimistic ``high'' configuration ($B_0 = 10^6$~G, $\dot{M} = 10^{-1}\dot{M}_{\rm Edd}$). When integrated over the full X-ray range, the total synchrotron luminosity can approach or exceed the thermal emission only at the highest masses and only when compared to the non-spinning ($a=0$) lower edge of the thermal band. However, the non-thermal synchrotron emission in our framework is most efficient for rapidly spinning black holes capable of sustaining strong magnetospheres, as the available Blandford--Znajek power increases with spin. Including spin-dependent corrections to the thermal radiative efficiency, as illustrated by the $a=0.99$ upper edge of the thermal band in Fig.~\ref{fig:compare_lx_sweeps}, raises the thermal luminosity and restores a clear separation between the thermal and non-thermal components across the full mass range, regardless of whether the synchrotron emission is integrated over the soft or full X-ray band. Thus, even under extreme magnetic field strengths and accretion rates, the reconnection-powered synchrotron emission is unlikely to mask the pronounced thermal X-ray drop expected in the final stages of MBHB inspiral. We also note that the thermal luminosities shown here correspond effectively to face-on viewing geometries, consistent with the spectra presented in \citet{Krauth2023}. Relativistic Doppler boosting at higher inclinations can enhance the high-energy thermal emission (see Appendix~B of \citet{Krauth2023}), which would further increase the separation between the thermal and non-thermal components.

\begin{figure*}
    \centering
    \includegraphics[width=0.95\textwidth]{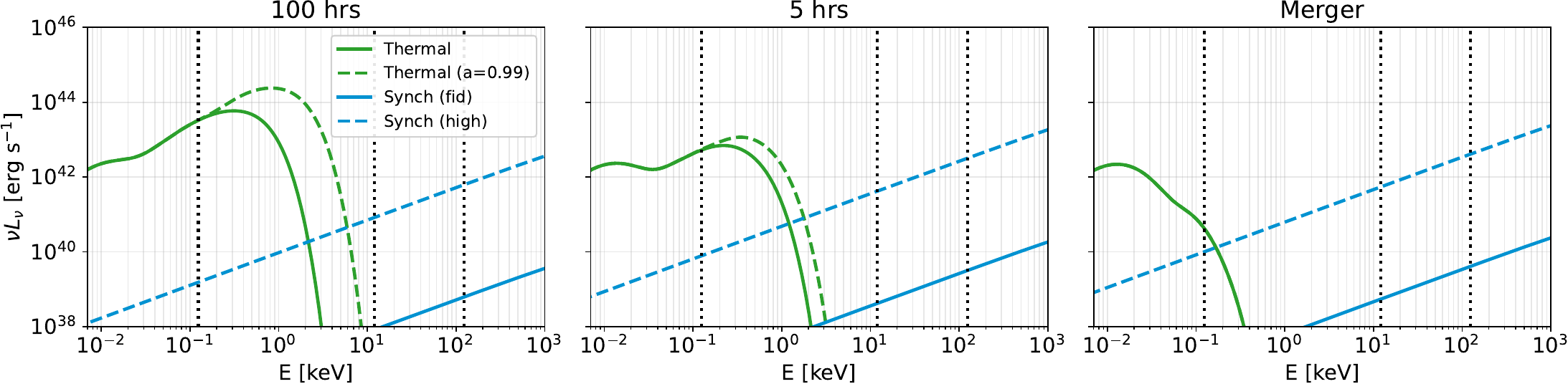}
    \caption{Spectral evolution of thermal and non-thermal emission during the final stages of MBHB inspiral. Shown are spectra at 100 hours before merger (left), 5 hours before merger (middle), and at merger (right). The peaked spectra (green) correspond to the thermal minidisc emission, with the solid curve showing the non-spinning case ($a=0$) and the dashed curve the maximally spinning case ($a=0.99$). The steadily rising power-law spectra (blue) show the non-thermal synchrotron emission powered by magnetic reconnection, where the solid line denotes the fiducial magnetic-field configuration and the dashed line represents the ``high'' magnetic-field model described in Appendix~\ref{app:synch_model}. Vertical dotted lines indicate representative soft and hard X-ray energy bands. At early times the spectrum is dominated by thermal emission, particularly in the soft X-ray band, while toward merger the thermal component diminishes and the synchrotron contribution becomes increasingly important at higher energies.
}
    \label{fig:spectral_evo}
\end{figure*}

In addition to comparing band-integrated luminosities, we examine the time-dependent spectral evolution of both the thermal and non-thermal components, allowing us to assess how their relative contributions evolve across soft and hard X-ray energies as merger approaches. Figure~\ref{fig:spectral_evo} shows representative spectra evaluated at 100 hours before merger, 5 hours before merger, and at merger. For each epoch, we plot the thermal spectrum for both non-spinning ($a=0$) and rapidly spinning ($a=0.99$) black holes, together with the non-thermal synchrotron spectra for the fiducial and ``high'' magnetic field models. Vertical grey lines indicate the full X-ray range, broken into soft and hard bands.

At early times (100 hours before merger), the thermal component dominates the spectrum, notably in the soft X-ray band. The maximally spinning thermal model exhibits a higher peak luminosity and modestly harder spectrum relative to the non-spinning case, reflecting the increased radiative efficiency and smaller effective inner disc radius. In contrast, the synchrotron emission remains comparatively suppressed at large binary separations. This behaviour is consistent with reconnection-powered models in which the non-thermal luminosity strengthens as the binary separation decreases and magnetic stresses intensify (e.g., \citealt{Manikantan2025}). Even in the ``high'' magnetic field configuration, the non-thermal component contributes primarily at higher photon energies at these early stages.

By 5 hours before merger, the thermal emission has declined substantially as the minidiscs become increasingly truncated and depleted. The spectral peak shifts downward in luminosity, and the soft X-ray band begins to dim appreciably. Over the same interval, the synchrotron spectrum strengthens and extends to higher energies, with the ``high'' model becoming comparable to the thermal emission at hard X-ray energies while remaining subdominant in the soft band.

At merger, the thermal component is strongly suppressed across the X-ray range as the minidiscs effectively vanish. In this phase, the non-thermal synchrotron spectrum dominates the high-energy X-ray band, particularly for the ``high'' magnetic field configuration, while the fiducial model provides a more moderate but still significant high-energy contribution.

For the maximally spinning ($a=0.99$) thermal model, spin-dependent thin-disc corrections are applied only during the earlier, radiatively efficient phase. In our framework, we model the late inspiral and merger stages as transitioning toward a magnetically arrested disk (MAD) state, in which a significant fraction of the accretion power is redirected into large-scale magnetic fields and jet production rather than radiated thermally. In this regime, the assumptions underlying thin-disc radiative efficiency are no longer strictly applicable. We therefore suppress spin-dependent thermal enhancements at late times to avoid artificially boosting the thermal luminosity in a magnetically dominated phase where non-thermal processes are expected to play an increasingly important energetic role.

To connect the evolving spectra to observational diagnostics more directly, we next compute the time-dependent luminosity in representative instrument-specific X-ray bands. This allows us to assess how the relative thermal and non-thermal contributions evolve within the energy ranges probed by current and future facilities.

\begin{figure*}
    \centering
    \includegraphics[width=0.95\textwidth]{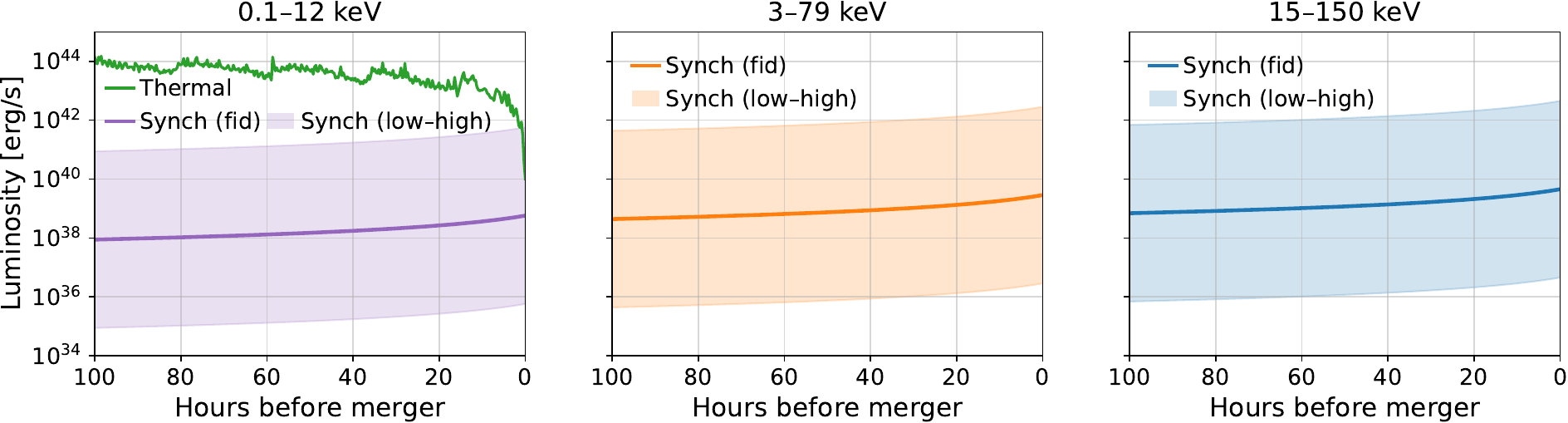}
    \caption{Time-dependent X-ray luminosity during the final $\sim100$ hours before merger in representative instrumental bands. The left panel shows the soft X-ray band (0.1--12\,keV), while the middle and right panels show the hard X-ray bands 3--79\,keV and 15--150\,keV, respectively. In the soft band, the solid curve denotes the thermal luminosity extracted from the hydrodynamical simulations. The non-thermal synchrotron contribution is shown in each panel by the lower curves: the solid line indicates the fiducial reconnection-powered model, while the surrounding shaded region brackets the range spanned by the ``low'' and ``high'' magnetic-field configurations described in Appendix~\ref{app:synch_model}. Although the thermal emission dominates the soft X-ray band, it contributes negligibly at higher photon energies, leaving the hard X-ray bands largely governed by the non-thermal synchrotron component. As merger approaches, the thermal luminosity drops sharply while the synchrotron contribution gradually strengthens.
}
    \label{fig:band_lcs}
\end{figure*}

Figure~\ref{fig:band_lcs} shows the evolution of the band-integrated luminosity over the final 100 hours before merger across three representative X-ray energy bands. The left panel shows the soft X-ray band ($0.1$--$12$ keV), relevant for current and forthcoming soft X-ray facilities such as \textit{Chandra}, \textit{XMM-Newton}, \textit{XRISM}, \textit{eROSITA}, and \textit{Athena} mission \citep{Weisskopf2002,Jansen2001,Tashiro2021,Predehl2021,Nandra2013}. In this band the thermal emission dominates early and drops sharply as the minidiscs are depleted during the late inspiral. For fiducial parameters the synchrotron contribution remains several orders of magnitude lower throughout the entire inspiral, becoming comparable only near merger in optimistic configurations.

The middle and right panels show the harder X-ray bands probed by \textit{NuSTAR} ($3$--$79$ keV) and \textit{Swift}-BAT ($15$--$150$ keV), respectively \citep{Harrison2013,Barthelmy2005}. Because the thermal spectrum is effectively confined to photon energies below $\sim12$~keV, its contribution to these higher-energy bands is negligible, leaving the emission largely governed by the non-thermal synchrotron component. Toward merger, this non-thermal luminosity gradually increases. While the strength of the hard-band enhancement depends sensitively on magnetic-field strength and accretion rate, its emergence near merger reflects the increasing dynamical importance of magnetic stresses as the system transitions toward a magnetically dominated state.

Taken together, these band-resolved light curves indicate that the final stages of MBHB inspiral may exhibit not only a luminosity decline, but a spectral reconfiguration, with soft X-rays tracing minidisc depletion and hard X-rays probing the growing influence of magnetospheric reconnection. While this enhancement is most pronounced in the hard X-ray bands and its observational detectability depends sensitively on source mass, distance, and exposure time, the resulting spectral transition may nevertheless provide an observational handle on the approach to merger, linking the soft X-ray disappearance with a potential hard-band signature of magnetospheric activity.

\begin{figure*}
    \centering
    \includegraphics[width=0.9\textwidth]{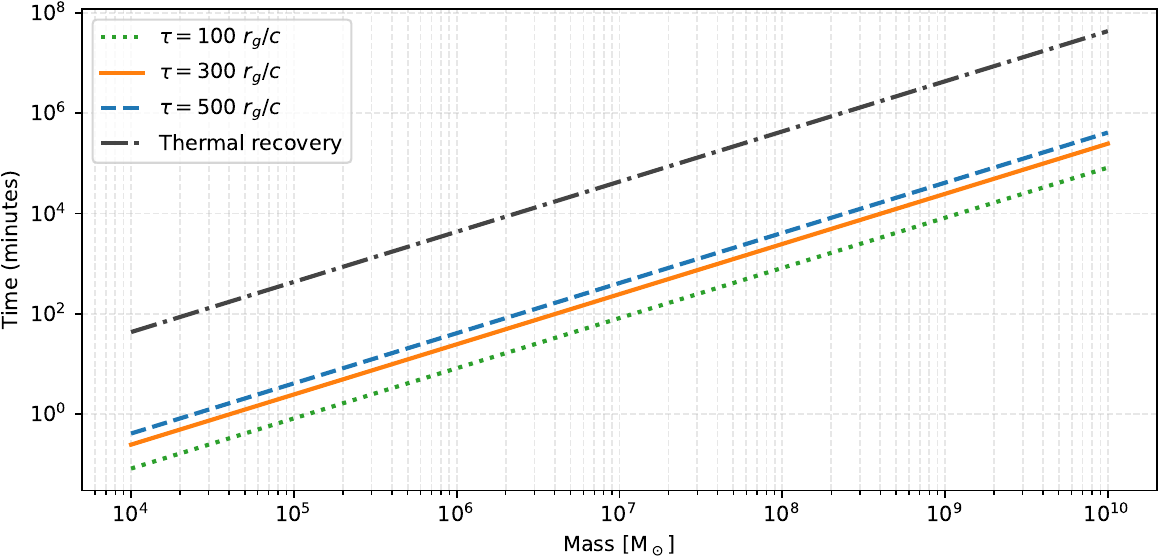}
    \caption{Time required for the non-thermal X-ray luminosity to decay by one e-fold due to magnetospheric balding, shown as a function of black hole mass. The three lower curves correspond, from bottom to top, to decay timescales $\tau = 100\,r_g/c$ (dotted), $300\,r_g/c$ (solid), and $500\,r_g/c$ (dashed). The upper dash–dotted curve shows an illustrative recovery timescale for the thermal X-ray emission, defined as the time required for the thermal luminosity to return to within two orders of magnitude of its pre-merger level. Across the full MBHB mass range, the non-thermal decay timescale remains shorter than the corresponding thermal recovery timescale, implying that any transient non-thermal enhancement fades before the thermal emission is re-established.
    }
    \label{fig:balding_efold}
\end{figure*}

The duration of this high-energy phase is constrained by the subsequent evolution of the magnetosphere. Figure~\ref{fig:balding_efold} shows the characteristic timescale on which reconnection-powered non-thermal X-ray emission is expected to decay once the external magnetic flux supply to the black hole magnetosphere is disrupted. Following the magnetospheric balding framework of \citet{Bransgrove2021}, the horizon-threading magnetic flux dissipates quasi-exponentially with an e-folding time $\tau \sim 100$--$500\, r_g/c$. Because this gravitational timescale scales linearly with black hole mass, the corresponding decay time increases proportionally across the MBHB mass range, from minutes for $M \sim 10^{4-5}\,M_\odot$ to days for $M \sim 10^{9-10}\,M_\odot$. This decay sets the duration of any merger-associated hard X-ray enhancement.

Figure~\ref{fig:balding_efold} also shows an illustrative recovery timescale for the thermal X-ray emission following merger. Although the precise definition of “recovery” is not unique, this curve marks the time required for the thermal luminosity to return to within two orders of magnitude of its pre-merger level—comparable to even the most optimistic non-thermal contribution considered here. Across the full mass range, this thermal recovery time exceeds the reconnection-driven decay time. Consequently, even in optimistic magnetic-field configurations where the non-thermal component briefly approaches the declining thermal luminosity at high photon energies, the reconnection-powered emission fades before the thermal emission rebounds. The system is therefore expected to enter a transient X-ray–dark phase following merger, until accretion and magnetic flux are re-established.

As shown in Fig.~\ref{fig:compare_lx_sweeps}, only under the most extreme magnetic field strengths and accretion rates does the reconnection-powered synchrotron emission begin to compete with the thermal X-ray luminosity, and then primarily in the full X-ray band at the highest black hole masses. Reaching such levels in our framework requires magnetospheric field strengths significantly above those typically expected for magnetically arrested accretion flows and accretion rates approaching or exceeding the Eddington limit—conditions that are both optimistic and likely short-lived during the late inspiral phase when the minidiscs are simultaneously being depleted. Even in this regime, the balding-driven decay timescales shown in Fig.~\ref{fig:balding_efold} imply that any transient non-thermal enhancement would fade rapidly once the magnetic flux supply is disrupted. Thus, while extreme configurations may briefly elevate the hard X-ray output, they do not prevent the system from entering the merger-associated X-ray–dark phase described above.

\section{Conclusions and Discussion}
\label{sec:Con}

In this work, we have investigated whether non-thermal X-ray emission powered by magnetic reconnection could obscure or modify the recently identified disappearing thermal X-ray signature associated with merging massive black hole binaries embedded in circumbinary disks. Motivated by hydrodynamical simulations showing a sharp decline in thermal X-ray emission from tidally truncated minidiscs near merger \citep{Krauth2023}, we constructed semi-analytic models for both the thermal and reconnection-powered non-thermal components. We compared their mass scalings, spectral shapes, and time-dependent evolution across soft and hard X-ray bands over the LISA-relevant MBHB mass range, and incorporated a magnetospheric balding prescription to model the post-merger decay of the non-thermal emission.

Our primary result is that, for physically motivated magnetic field strengths and accretion rates, the non-thermal synchrotron X-ray luminosity remains subdominant to the thermal X-ray emission throughout the inspiral. In the soft X-ray band—where the thermal emission is concentrated and where most X-ray instruments operate—the synchrotron component remains below the thermal level even under optimistic assumptions. When integrated over the full X-ray range, the non-thermal luminosity can approach or exceed the thermal component only at the highest black hole masses and only when compared to a non-spinning ($a=0$) thermal model. For rapidly spinning black holes ($a=0.99$), which are expected to produce the strongest magnetospheres and thus the largest synchrotron output in our framework, the thermal emission remains dominant across the full mass range. This indicates that the disappearance of the thermal X-ray signal identified by \citet{Krauth2023} is unlikely to be masked by reconnection-powered synchrotron radiation in realistic astrophysical systems.

Related non-thermal emission channels may also operate in merging MBHB systems, particularly those associated with jet activity on larger spatial scales. For example, models of magnetic reconnection in dual jet--jet interaction regions predict non-thermal radiation originating at distances well beyond the near-horizon magnetosphere \citep{Gutierrez2024}. In such scenarios, the resulting synchrotron emission is spatially extended, varies slowly, and peaks primarily at infrared to optical/UV energies, with only a modest contribution to the X-ray band even for optimistic acceleration efficiencies. Such emission would therefore act as a quasi-steady luminosity pedestal rather than a rapidly evolving signal tied to the near-merger depletion of the minidiscs. As a result, while jet--jet interactions may represent an additional non-thermal channel in MBHB systems, they are unlikely to mask or mimic the compact, merger-associated thermal X-ray drop that is the focus of this work.

In addition to band-integrated comparisons, our spectral modelling shows that the relative dominance of thermal and non-thermal emission is energy-dependent and evolves across inspiral. At early times the soft X-ray band is clearly thermal-dominated, while toward merger the thermal component collapses and the non-thermal synchrotron emission becomes increasingly important at higher photon energies. However, this transition occurs only after the thermal luminosity has already declined substantially, and the non-thermal component remains insufficient to erase the characteristic soft X-ray disappearance identified by \citet{Krauth2023}.

The spectral reconfiguration towards harder photon energies near merger therefore points to a complementary high-energy signature rather than a competing one. A transient enhancement in the hard X-ray band could trace the increasing role of magnetospheric reconnection during the final stages of inspiral. If detectable in nearby systems with favourable combinations of mass, magnetic field strength, and accretion state, such hard-band emission would not obscure the thermal-drop scenario, but instead mark the system’s transition toward a magnetically dominated state, providing an additional observational handle on the approach to merger.

We further considered the temporal evolution of the non-thermal emission by incorporating the magnetospheric balding framework of \citet{Bransgrove2021}. Once the external supply of magnetic flux diminishes near merger, reconnection rapidly dissipates the horizon-threading magnetic field, causing the non-thermal emission to decay quasi-exponentially on timescales of $\tau \sim 100$–$500\,r_g/c$. Although this decay time increases linearly with black hole mass, the same mass scaling applies to the duration of the thermal X-ray–dark phase found in hydrodynamical simulations, and to the subsequent recovery of thermal X-ray emission. As a result, even in scenarios where non-thermal emission approaches the thermal level prior to merger, the reconnection-powered component is expected to fade more rapidly than the thermal emission recovers, leaving the system temporarily X-ray dark until accretion and magnetic flux are re-established.

Taken together, these results reinforce the disappearing thermal X-ray signature as a robust electromagnetic counterpart to MBHB mergers. Because the reconnection-powered component does not persist at a level capable of masking the thermal drop, the merger-associated X-ray–dark phase remains a distinct and temporally localized feature suitable for identifying host galaxies within the LISA localization volume and performing multi-messenger follow-up observations. Any accompanying hard X-ray enhancement, if detected, would not contradict this picture, but instead provide complementary information about the onset of magnetospheric dominance marking the final stages of inspiral. In combination, the soft-band disappearance and possible hard-band emergence could therefore offer a multi-band electromagnetic framework for temporally associating LISA detections with their astrophysical environments and advancing broader multi-messenger science goals.

Several caveats remain. Our non-thermal model is necessarily idealized and calibrated to kinetic reconnection simulations in simplified geometries, while the true magnetospheric structure of merging MBHBs may be more complex. Magnetic reconnection may also produce inverse-Compton emission, as explored in radiative PIC simulations by \citet{Hakobyan2023}. Toward lower black hole masses, the magnetization increases strongly, while the characteristic synchrotron-limited Lorentz factor decreases due to enhanced radiative cooling; at the same time, the characteristic energy of thermal seed photons shifts to higher frequencies for
lower-mass systems. The combination of harder seed photons and relativistic pairs
therefore places inverse-Compton scattering deep in the Klein--Nishina regime at the
low-mass end of our sweep, strongly suppressing IC relative to synchrotron. Efficient IC emission is expected primarily for the highest-mass systems, while radiative PIC simulations further reveal an intermediate spectral excess associated with
intermittent plasmoid dynamics. Quantifying the importance of this feature across
merger requires fully self-consistent simulations, and we therefore treat any
inverse-Compton contribution as a secondary, model-dependent effect. In addition, highly asymmetric systems, extreme spin configurations, or transient episodes of enhanced magnetic flux accumulation could in principle modify the relative balance between thermal and non-thermal emission. We also note that recent 3D GRMHD simulations have shown that, in certain regions of parameter space, accretion onto the binary may persist through merger \citep{Ennoggi2025}, potentially weakening or delaying the thermal X-ray drop. Future global GRMHD simulations that self-consistently capture both minidisc disruption and magnetospheric reconnection across merger will be essential for fully quantifying these effects.

Nevertheless, within the range of parameters explored here, our results indicate that magnetic reconnection is unlikely to erase or substantially alter the disappearing thermal X-ray signature. The spectral evolution further demonstrates that while non-thermal emission can become prominent at higher photon energies near merger—particularly in extreme magnetic-field configurations—the soft X-ray disappearance associated with minidisc depletion remains robust. Although the non-thermal component may briefly rise as magnetic stresses intensify, it subsequently decays on short balding timescales once the external magnetic flux supply diminishes. As a result, the system is expected to enter a transient X-ray–dark phase before accretion and magnetic flux are re-established. This behaviour underscores X-ray observations as a powerful complement to gravitational-wave detections in probing the final stages of massive black hole binary evolution.

\section*{Acknowledgements}

We acknowledge support from the NWA Roadmap grant ``GW LISA/ET: Shivers from the Deep Universe: a National Infrastructure for Gravitational Wave Research'' (LK), JD is supported by NASA through the NASA Hubble Fellowship grant HST-HF2-51552.001A, awarded by the Space Telescope Science Institute, which is operated by the Association of Universities for Research in Astronomy, Incorporated, under NASA contract NAS5-26555.
{\it Software:} {\tt python} \citep{travis2007,jarrod2011}, {\tt scipy} \citep{jones2001}, {\tt numpy} \citep{walt2011}, and {\tt matplotlib} \citep{hunter2007}.

%%%%%%%%%%%%%%%%%%%%%%%%%%%%%%%%%%%%%%%%%%%%%%%%%%
\section*{Data Availability}
The data underlying this article will be shared on reasonable request to the corresponding author.

%%%%%%%%%%%%%%%%%%%%%%%%%%%%%%%%%%%%%%%%%%%%%%%%%%%
%%% ______________________________________________

\bibliographystyle{mnras}

\appendix
\section{Fiducial Non-thermal Synchrotron Model}
\label{app:synch_model}

This appendix summarizes the equations and fiducial parameters used to compute the reconnection-powered non-thermal X-ray emission discussed in Section~2.2.

The near-horizon magnetic field strength is parameterized as
\begin{equation}
B(M,\dot M) = B_0
\left(\frac{M}{10^6\,M_\odot}\right)^{-1/2}
\left(\frac{\dot M}{0.01\,\dot M_{\rm Edd}}\right)^{1/2}
\label{eq:A1},
\end{equation}
where $B_0$ is a normalization constant. We adopt a fiducial value $B_0 = 10^5\,{\rm G}$, corresponding to an upper-end realization of reconnection-based magnetospheric models when scaled to MBHB masses and accretion rates. This choice intentionally maximizes the non-thermal synchrotron component, making it more likely to compete with the thermal X-ray emission.

Given $B(M,\dot M)$, the jet power is taken to scale as
\begin{equation}
L_{\rm BZ} \propto B^2 M^2 ,
\end{equation}
with spin-dependent factors absorbed into the overall normalization. A fixed fraction of this Blandford--Znajek jet power is assumed to be dissipated and radiated via magnetic reconnection,
\begin{equation}
L_{\rm rec} = f_{\rm rec} L_{\rm BZ},
\end{equation}
where we adopt $f_{\rm rec} = 0.1$ as a fiducial reconnection efficiency.

The synchrotron spectrum is computed assuming a non-thermal particle distribution
\begin{equation}
N(\gamma) \propto \gamma^{-p}\exp(-\gamma/\gamma_c),
\end{equation}
with $p=1.25$ and $\gamma_c=5\times10^{7}$. These values are chosen to match kinetic simulations of radiative relativistic magnetic reconnection, which produce hard particle spectra with $1\lesssim p\lesssim1.5$ and cutoff Lorentz factors $\gamma_c\sim10^{7}$--$10^{8}$ set by the synchrotron burnoff limit \citep{Hakobyan2023}.

\label{lastpage}
\end{document}